\documentclass[12pt,preprint]{aastex}

\slugcomment{Accepted by The Astrophysical Journal}

\shorttitle{CCS and NH$_{3}$ associated with low-mass YSOs}
\shortauthors{de Gregorio-Monsalvo et al.}
      
\begin{document}


\title{CCS and NH$_{3}$ emission associated with low-mass young stellar objects}


\author{Itziar de Gregorio-Monsalvo\altaffilmark{1}, Jos\'e
  F. G\'omez\altaffilmark{1,2}, Olga Su\'arez\altaffilmark{1}, 
   Thomas B. H. Kuiper\altaffilmark{3}, Luis
   F. Rodr\'{\i}guez\altaffilmark{4}, Elena Jim{\'e}nez-Bail{\'o}n\altaffilmark{5}} 

\altaffiltext{1}{Laboratorio de Astrof\'{\i}sica Espacial y F\'{\i}sica
  Fundamental (INTA), Apartado 50727, E-28080 Madrid, Spain}
\altaffiltext{2}{Instituto de Astrof\'{\i}sica de Andaluc\'{\i}a
  (CSIC), Apartado 3004, E-18080 Granada, Spain}
\altaffiltext{3}{ Jet Propulsion Laboratory, California Institute of Technology, USA}
\altaffiltext{4}{Centro de Radioastronom\'{\i}a y Astrof\'{\i}sica, UNAM,
  Apartado Postal 3-72 (Xangari), 58089 Morelia, Michoac\'an. Mexico}
\altaffiltext{5}{XMM-Newton Science Operation Center/RSSD-ESA,
  Apartado 50727, E-28080 Madrid, Spain.}



\begin{abstract}
In this work we present a sensitive and systematic single-dish survey
of CCS emission (complemented with  ammonia observations) 
at 1 cm, toward a sample of low- and intermediate-mass young star
forming regions known to harbor water maser emission, made
with NASA's 70 m antenna at Robledo de Chavela, Spain.
  Out of the 40 star forming
regions surveyed in the CCS(2$_{1}$$-$1$_{0}$) line, only 6 low-mass 
sources
show CCS emission: one transitional object between pre-stellar and
protostellar Class 0 phase (GF9-2), three Class 0 protostars 
(L1448-IRS3, L1448C, and B1-IRS), a Class I 
source (L1251A), and a young T Tauri star (NGC2071-North). 
 Since CCS is considered an ``early-time'' ($\la
10^5$ yr)
molecule, we explain these results by either proposing a revision of the
classification of the age of NGC2071-North and L1251A, or suggesting
the possibility that
the particular physical conditions and processes of each source affect the
destruction/production of the CCS. No
statistically significant relationship was found between the presence of CCS
and parameters of the molecular outflows and their driving
sources. Nevertheless, we found a significant relationship
between the detectability of CCS and the ammonia peak intensity
(higher in regions with CCS), but  not
with its integrated intensity. 
This tendency found may suggest that the narrower ammonia line widths
in the less turbulent medium associated with younger cores may
compensate for the
differences in ammonia peak intensity, rendering differences in integrated
intensity negligible.
From the CCS detection rate we derive a lifetime of this molecule of $\simeq$
(0.7$-$3)$\times$10$^{4}$ yr in low-mass star forming regions. 

\end{abstract}


-- 

\section{Introduction}
The lines of the CCS molecule are a powerful tool for studying the physical
conditions and the structure of dark molecular clouds, because they are
intense, abundant and not very opaque in these regions. In addition,
 CCS is very useful for performing dynamical studies, since it is heavier 
than other high density gas tracers and it has no hyperfine 
structure \citep{Sai87,Suz92}.

Moreover, CCS is useful for obtaining information about the
age of molecular clouds. Previous single-dish observations show that
CCS lines are intense in starless, cold, quiescent cores, while ammonia
tends to be abundant in star forming regions \citep{Suz92}. 
It has been suggested that CCS is present in the first stages of
molecular cloud evolution, but it is soon destroyed (on a timescale of
$\simeq$10$^{5}$ years, \citealt{Mil90,Suz92}) after the formation of a
dense core. This destruction process is induced by the core
contraction \citep{Suz92} that initiates star formation. 
On the other hand, when the molecular cloud
evolves, the physical conditions and the chemical evolution in
molecular cores favor the formation of other molecules like 
NH$_{3}$ \citep{Suz92}.
 For this reason the abundance ratio [NH$_{3}$]/[CCS] has been
considered as an indicator of the evolution of molecular cores. 
This time-dependent chemistry is responsible for the pronounced spatial
anticorrelation observed in the emission from these two species, 
where ammonia tends to trace the inner regions and the CCS
is found to be located outside, surrounding ammonia cores, 
in a clumpy distribution \citep{Hir92, Kui96, Lai03}.

If this evolutionary trend (CCS more abundant in starless cores and 
NH$_{3}$ in more evolved, active star forming regions) is correct, 
any chance to use
CCS as a tool to study star-formation processes will necessarily
happen during the very first stages of the stellar evolution, i.e.,
Class 0 protostars.
Interestingly, in these young protostars some of the key phenomena
characteristic of the star formation (infall, disk formation and
powerful mass-loss phenomena) are especially prominent and coeval, and 
have their own kinematical signature, which make 
kinematical studies of their environment especially interesting.
In this early stage of stellar evolution, the
presence of water maser emission at 22 GHz is rather common. This emission 
 is considered a good tracer of mass-loss activity in young stellar
 objects (YSOs) in general \citep{Rod80,Deb05}, and a good indicator of the age of low-mass YSOs, since they tend to be excited preferentially in the 
Class 0 stage \citep{Fur01}.

The critical scales for these star formation processes are of the
order of $\simeq 100-1000$ AU in the case of low mass stars ($\leq
1\arcsec-10\arcsec$ for nearby molecular clouds; 
\citealt{Bel01,Che95,Cla98}).
In order to study these phenomena with high enough resolution, it is
necessary to carry out interferometric
observations. This is feasible for strong maser lines, but
very difficult for thermal emission (like CCS and ammonia) since it
is usually weak and therefore requiring 
high sensitivity is needed.
At wavelengths of $\sim 1$ cm an instrument like the Very Large Array 
(VLA) provides
high angular resolution, but at the expense of worse atmospheric
phases that blur the emission. Weak sources, in which self-calibration
is not possible, would not be detected, or the quality of their maps
 would be poor. Nevertheless, these kinds of problems have been addressed,
in the case of continuum emission, by using the cross-calibration
technique. This method involves the simultaneous observation of 1 cm
continuum and strong water masers, and the use of the self-calibration
solutions of the latter to correct phase and amplitude errors in the
weaker continuum \citep{Tor96, Tor97, Tor98}. This technique of
cross-calibration using water masers has never been applied to
spectral lines.
The CCS(2$_{1}$$-$1$_{0}$)  transition at $\sim$22 GHz is probably
the best
candidate to perform it, since it is only 110 MHz away from the H$_{2}$O 
(6$_{16}$$-$5$_{23}$) maser line.

A high-resolution study of the kinematics and the chemical evolution 
of the environment of the 
Class 0 source B1-IRS was carried out  by \citet{deG05}.
In that work, CCS, ammonia, and water maser emissions at 1 cm
were observed with the VLA, although
cross-calibration was not possible there, since the maser 
was not strong enough.
Three different CCS clumps were detected,
 whose kinematical pattern was interpreted as gas interacting with the molecular outflow that exists in
the region. This interaction was not observed using other molecular
tracers or with lower-resolution observations of CCS in
this source \citep{Hir02,Lai03}. The evidence of interaction with the
outflow led \citet{deG05} to suggest the possibility that CCS
abundance could be enhanced via shock-induced chemistry. Moreover, in
that region  a spatial anticorrelation between CCS and ammonia at
scales of $\simeq$5$\arcsec$ was observed for the first time, which
illustrates the importance of time-dependent chemistry on small
spatial scales.

In this work we present a sensitive and systematic single-dish survey
of CCS emission (complemented with NH$_{3}$
observations) at 1 cm wavelength, toward low-mass star forming regions 
that are known to harbor water maser emission, using the NASA's 70m antenna
at Robledo de Chavela, Spain.
One of our main aims is to find the best candidates to make
interferometric CCS studies as the one made in B1-IRS with the VLA,
but applying the cross-calibration technique for obtaining high-quality
maps. Interferometric maps of CCS, ammonia and water masers would
allow us to study kinematical and physical properties of star forming 
regions at high resolution.    
Another purpose of the present survey is the search for the youngest
star forming regions, assuming that the presence of 
water masers and CCS emission are signs of youth in star forming
regions. 
Moreover, we present a search for the relation between the CCS emission and the physical 
characteristics of the star forming regions of the survey, 
focusing on the molecular outflow properties, considering the possible
association between CCS emission and the outflow suggested for
 B1-IRS.

This paper is structured as follows: in section \S 2 we describe our
observations and data reduction. In section \S 3 we show the survey
results, as well as a short description of the sources detected. We
discuss the results in \S 4 and, finally, we summarize our conclusions in \S 5.

\section{Observations}
Observations were carried out in different periods between 2002 April
and 2005 March, using the NASA's 70-m antenna (DSS-63) at Robledo de
Chavela, Spain. The total telescope time was $\simeq$ 50 hours (without 
including calibration or pointing checks). We observed the CCS J$_{N}$=2$_{1}$-1$_{0}$ transition
(rest frequency $=$ 22344.033 MHz) and the NH$_{3}$(1,1) inversion
transition (rest frequency = 23694.496 MHz) toward a sample of star
forming regions.     
This antenna has a 1.3 cm receiver using cooled high electron
mobility transistor (HEMT) amplifiers. Calibration was performed using
a noise diode. The half power beam width at these frequencies is
$\simeq 40\arcsec$, and the mean beam efficiency is $\simeq 0.4$. 

The sample surveyed (Table~\ref{tbl-source}) consists of 40 low-mass
($L_{\rm bol} < 100$ L$_{\odot}$) and
intermediate-mass ($100 < L_{\rm bol} < 10^4$ L$_{\odot}$) 
star forming regions associated with water maser
emission, as a sign of youth and ongoing star formation processes.
Most of the sources are low-mass star forming regions, although we
have included 12 intermediate-luminosity star forming regions as a 
control sample, in order to test any possible relation between the 
CCS emission and
the mass of the sources.   
Observations were centered at positions where water maser emission has
been reported in the literature.  

Most of the CCS observations were performed in frequency-switching
mode. We used a 256 channel digital autocorrelator spectrometer with a
bandwidth of 1 MHz, centered at the $V_{LSR}$ of each cloud 
(see Table~\ref{tbl-source}), which provides a velocity resolution of 0.05
km s$^{-1}$. Seven sources (see Table~\ref{tbl-CCS}) were observed in
position-switching mode, with a 384 channel digital autocorrelator
spectrometer with a bandwidth of 2 MHz (velocity resolution of 0.07 km
s$^{-1}$). The system temperature varied from 44 to 135 K, depending
on weather conditions and elevation, with a mean system temperature of 76 K.

The ammonia observations were also made in position-switching mode, using the
384 channel digital autocorrelator spectrometer. We used a bandwidth of
16 MHz, centered at the $V_{LSR}$ of each cloud, which provides a
velocity resolution of 0.5 km s$^{-1}$. NGC 2071-North and AFGL 5157 were observed using a bandwidth 
of 4 MHz, with a velocity resolution of 0.13 km s$^{-1}$,
but the spectra were smoothed to a final resolution of 0.5 km
s$^{-1}$. The system
temperature varied from 52 to 99 K,  with a mean system temperature of 78 K.

The rms pointing accuracy was better than 10$\arcsec$.
The data reduction was performed using the CLASS package, which is
part of the GAG software package developed at IRAM and the
Observatoire de Grenoble.

\section{Survey Results}
\begin{figure}
\epsscale{0.9}
\plotone{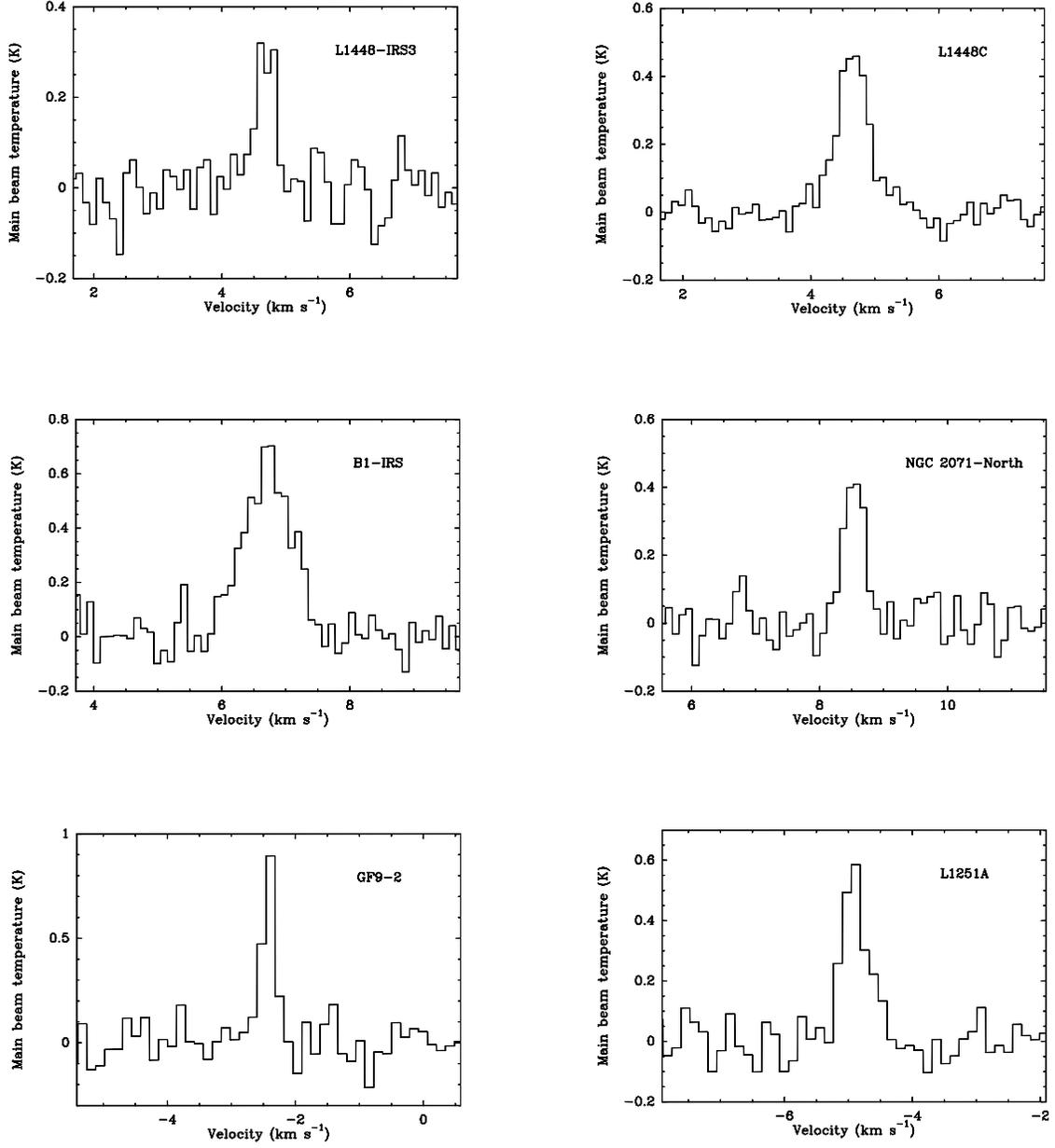}

\caption{Spectra of CCS(2$_{1}$-1$_{0}$) transition detected with the Robledo 70m antenna.}
\label{Spectra}
\end{figure}

In Table~\ref{tbl-CCS} and Table~\ref{tbl-NH3} we summarize the results of the survey. 
Out of the 40 star forming regions surveyed in the
CCS (2$_{1}$-1$_{0}$) line, only 6 low-mass sources show emission, and their
spectra are shown in Fig~\ref{Spectra}. We have not detected CCS
emission towards any of the 12 intermediate-mass objects.
On the other hand, we also observed 36 of the sources in the NH$_{3}$ (1,1)
transition, of which 31 show detectable emission. 
We note that all sources detected in CCS are associated with ammonia
emission.
In order to compare our results with the work by \citet{Suz92}, who
surveyed dark cores mostly in Taurus and Ophiucus, we will 
consider only the low-mass sources in our sample. 
We have excluded the source
L260 from statistical calculations, since its noise level for CCS is
significantly higher than that of the other sources. In fact, several of our
CCS detections are below the upper limit given for L260 (Table
\ref{tbl-CCS}). 
With this exclusion, we have a subset of 27 low-mass sources, among which 
we can strictly define as sources without
detectable CCS emission those with $T_{\rm MB} < 0.3$ K or $N_{\rm
  CCS} < 2.1\times 10^{12}$ cm$^{-3}$. 
Taking from \citet{Suz92}
the star forming regions (considering as such those with presence of
outflows and/or IRAS sources), the detection rates obtained for CCS is 23\%
 at 22 GHz. Since their detection limit of the CCS column density is similar to ours, 
we can meaningfully compare it
with our detection rate of 22\% for CCS. Using Fisher's exact test, we determined that 
our detection rates for CCS is statistically
compatible  with that of
\citet{Suz92}, at a confidence level of 95\%.\footnote{We have used
  this confidence level for all the statistical tests in this paper}
This compatibility is reasonable, 
since water maser emission (the selection criterion for our survey) is a
good tracer of star formation activity \citep{Fur03,Rod80}. 

Most of the star forming regions we surveyed have detectable ammonia
emission, which suggests that ammonia survives well in the protostar 
phase, as indicated by the presence of water masers. Few of the
regions have CCS emission, but all of these also have NH$_{3}$.
Therefore, there are no regions associated with both H$_{2}$O 
maser and CCS emission, but without NH$_{3}$. 
Bearing in mind that there are many CCS-emitting clouds without any
sign of harboring protostars \citep{Benson98, Suz92}, these results support the evolutionary sequence of
CCS+NH$_{3}$+H$_{2}$O $\rightarrow$ NH$_{3}$+H$_{2}$O. In particular,
it supports the idea that the CCS molecule is destroyed before ammonia.

\subsection{Notes on the sources detected in CCS}

\subsubsection{L1448 Region}
L1448 is a dense globule located in Perseus. This cloud contains
numerous signatures of mass-loss activity \citep{Bac90,Bac95b, Cur99} and it 
is an interesting environment for studying the possibility that outflows
from YSOs may induce new star formation in other regions of the
parental cloud.
Single-dish ammonia observations  reveal two
peaks located at the center and at the north-east of the cloud,
associated with L1448C and L1448-IRS3 respectively \citep{Ang89,Bac90}.   

L1448C \citep{Cur90} is catalogued as a Class 0 source
\citep{And93,Bar94} that drives
one of the most energetic and highly collimated molecular outflows ever
seen in a low-mass star forming region \citep{Bac90}. 
This outflow reveals a strong interaction with the northern material of
the cloud, associated with L1448-IRS3, the brightest
far-infrared source of the region. 

 L1448-IRS3 is known to consist of three individual Class 0 sources: a
 close proto-binary system [L1448N(A) and L1448N(B)], in which each
 component powers its own outflow, and a third source, L1448 NW \citep{Bar98,Cur90,Ter97}.
\citet{Bar98} suggested that the outflow powered by L1448C was responsible for   
the formation of the proto-binary system, whose
outflows may, in their turn, have produced the NW source. 

We detected
CCS emission towards both L1448-IRS3 and L1448C. This is the first
detection of emission from this molecule reported for either of them.

\subsubsection{B1-IRS}

B1-IRS (IRAS 03301+3057) is a far-infrared source located in the
Perseus OB2 complex \citep{Bac90b}. This source is classified as a
Class 0 source \citep{Hir97}, and it is associated with a CO molecular
outflow that displays a strong blueshifted emission \citep{Bac90b, Hir97}.

CCS emission at 22 GHz was detected for the first time in this region
by \citet{Suz92}. Later, \citet{Lai00} observed CCS at 33.8 GHz with
the BIMA interferometer, and their maps showed 
a clumpy distribution surrounding B1-IRS. 
This clumpy distribution was confirmed with VLA CCS observations at 22
GHz \citep{deG05}. The clumps exhibit a clear velocity gradient, 
interpreted as gas interacting with the molecular outflow. Moreover a CCS
local abundance enhancement via shock-induced chemistry was 
also suggested for the first time in that work.
B1-IRS was the first region where a spatial anticorrelation between
CCS and ammonia was detected at high angular resolution ($\simeq5\arcsec$,
\citealt{deG05}).

\subsubsection{NGC2071-North}
NGC2071 North is located at 20$\arcmin$ north of the reflection nebula NGC2071 \citep{Fuk86}.  
This region hosts a CO molecular outflow that lies near the source IRAS
05451+0037. Maps of the outflow at different angular
resolution display very different geometries:  a
bent-U shaped outflow at $17''$ resolution, which suggests strong interaction with the ambient
material \citep{Iwa88}, and overlapping blue- and redshifted emission
at  $\ga 50''$ resolution \citep{Iwa88,Gol92}, 
which is compatible with a bipolar outflow 
whose axis lies very near
the plane of the sky. 
IRAS 05451+0037 is coincident with a CS and a NH$_{3}$ peak
\citep{Gol92}, and it shows the
far-infrared spectrum  of a T Tauri star \citep{Iwa88}.
This source was suggested to be the central engine of the molecular
outflow \citep{Iwa88}, although this is somewhat uncertain due to
the complicated geometry of the outflow.

\citet{Suz92} detected CCS at 45 GHz toward this source. 
In our survey, we report CCS emission at 22 GHz for the first time.

\subsubsection{GF9-2}
This source is located in the filamentary quiescent dark cloud GF-9 (LDN 1082),
without any associated IRAS point sources or radio continuum emission
\citep{Cia98,Wie99}. 
It was catalogued  as a transitional object between pre-stellar
and proto-stellar Class 0 phase \citep{Wie99},
 which would turn GF-9 to be the youngest source in our survey. 
\citet{Fur03} detected the presence of weak redshifted CO wing
emission, and a weak H$_{2}$O maser (0.3 Jy). These
authors pointed out that GF9-2 was the lowest-luminosity (0.3
$L_{\odot}$) object known to harbor H$_{2}$O masers.
We have detected both CCS and ammonia emission lines toward this source for the first
time. 

\subsubsection{L1251A}
L1251A (IRAS 22343+7501) is located at the northern part of L1251, a small elongated 
cloud located in Cepheus. This source has been classified as a Class I
object \citep{Mar97,Nik03} 
and appears to be powering  the extended CO outflow
seen in this region \citep{Sat89}, as well as an optical jet
\citep{Bal92}. Several radio continuum sources were discovered near
IRAS 22343+7501 \citep{Mee98}. Recent VLA centimeter observations, reveal  that
this IRAS source may consist of two protostellar objects, with 
spectral indices consistent with thermal emission, and either of them
could be the driving source of the CO outflow \citep{Bel01}.
We report CCS emission toward this source for the first time.

\section{Discussion}
\subsection{The Lifetime of CCS in Star Forming Regions}
The detection rate obtained in our survey for CCS (22$\%$ of the low-mass sources
 show CCS emission) is similar to the results of \citet{Suz92}
in star forming regions (23$\%$ detections).  
The low rate of detection of CCS emission in star forming regions seems to
indicate that this molecule is soon destroyed after the
processes associated with star formation begin. This result, together
with the presence of water maser emission, allows us to roughly
estimate the lifetime of CCS in low-mass
star forming regions. The association of a low-mass central source with
water maser emission typically lasts for one third of the duration of
the embedded state \citep{Wil94, Cla96}. 
This embedded phase is estimated to last for $\simeq (1-4)\times 10^{5}$ yr \citep{Wil89b, And94, Che95}. 
Therefore, the water maser emission lasts for $\simeq (0.3-1.3)\times10^{5}$ yr in low-mass star forming objects.
 If we assume that the regions of our sample are homogeneously distributed in age along the $\simeq (0.3-1.3)\times10^{5}$ yr in which
water maser emission is observed, our detection rate of 6 out of 27,
would indicate that the CCS emission lasts for $\simeq (0.7-3)\times 10^{4}$
yr after star formation started. 

\subsection{Evolutionary stage of the SFRs associated with CCS emission}
As we mentioned in the introduction, since the CCS molecule has been 
considered in previous works as a youth tracer of molecular cores, 
the star forming regions associated with this emission should contain
some of the youngest YSOs. This evolutionary trend for the CCS
chemistry must be valid as a general, qualitative trend when we
compare large groups of molecular cores, although it may not stand in
a quantitative way in the comparison of particular sources. In particular,
the results of our survey do not totally agree with the first statement,
since four of the sources that
host CCS emission are catalogued as young Class 0 sources (GF9-2, L1448-IRS3,
L1448C, B1-IRS), but L1251A and NGC2071-North are catalogued 
as more evolved YSOs. 
There are two possible explanations for these
results: either the classification of the evolutionary stage of
 L1251A and NGC2071-North should be revised, or the particular
 physical conditions and processes of each
source affect the production and destruction of CCS. 

NGC2071-North (IRAS 05451+0037) shows a far-infrared spectrum
of a T Tauri star \citep{Iwa88}. Nevertheless, there is
 no optical counterpart in the POSSM prints, and no near-infrared
emission in the 2MASS catalog at the position of IRAS 05451+0037, which
suggests a more embedded nature of this source. The complicated geometry of the
molecular outflow of the region (see section \S 3.1.3), makes it difficult to know reliably
 whether IRAS 05451+0037 is the real engine of the outflow.
In order to clarify the real geometry of the outflow,
the power source, and its evolutionary stage, more accurate and
extended observations of CO, together with submillimeter observations, are needed.
In particular, the study of possible peaks at submillimeter
frequencies (typical of Class 0 objects) and the
measurement of the relation L$_{smm}$/L$_{bol}$ \citep{And93},
are especially important to elucidate the possible existence of other embedded
sources in the region, and to determine their evolutionary stage.

In the case of L1251A (IRAS 22343+7501), it is associated with a nebulosity visible at J, H, and K
2MASS-bands, and it is visible in POSSM plates. This source 
has been classified as a Class I YSO \citep{Mar97,Nik03}.
 Nevertheless, \citet{Mee98} discovered several radio continuum sources in the region near the
IRAS source, candidates to YSOs, and the CCS emission could be
associated with any of these embedded sources. In this case, higher
resolution observations of CCS emission would be useful to study its 
association with those sources.

The second possible explanation would be to accept that, in some
cases, CCS could be associated with sources in a more evolved stage. 
This could be the case if CCS emission depends not only on age, but also
on the particular phenomena related to the star
formation of each cloud. 
This possibility was already suggested in the high resolution
CCS study of B1-IRS by \citet{deG05}, who proposed that the
 CCS emission could be enhanced in the region where the 
outflow interacts with the surrounding cloud medium. The outflow interaction
with fresh low-density gaseous material that exists around dense cores,
could squeeze this low-density gas to the higher densities needed
($\simeq$ 10$^{5}$ cm$^{-3}$) to form this molecule and excite its
line emission. This phenomenon could explain the detection of this
molecule in stages more evolved than Class 0 sources.
The possible dependence of CCS on parameters different from age is
discussed in next section.

\subsection{Search for dependencies of CCS emission on source and
  cloud parameters}

The abundance ratio [NH$_{3}$]/[CCS] was proposed by \cite{Suz92} to be a
 good indicator of cloud evolution. This conclusion was based on theoretical 
pseudo-time-dependent calculations of the fractional abundance of CCS and 
NH$_{3}$, under conditions of constant H$_{2}$ density and constant gas 
kinetic temperature. The variation of the fractional abundances in both 
molecules was explained as an effect of the evolution of the molecular clouds, 
with CCS more abundant in starless cores, and ammonia more abundant in regions 
with signs of star formation activity. Moreover, \cite{Suz92} pointed out that
 this chemical evolutionary track was insensitive to the density of 
H$_{2}$. 

Nevertheless,  when the column densities of both molecules 
(N$_{\rm NH_{3}}$ and N$_{\rm CCS}$) are compared among
different clouds, 
the proposed anticorrelation disappears \citep{Suz92}.     
In our case, if we order the sources in which we detected CCS 
by their ratio of derived column densities N$_{\rm NH_3}$/N$_{\rm CCS}$,
they are, from lowest to highest ratio: B1$-$IRS, L1448C, GF9$-$2, L1448$-$IRS3, L1251A, and NGC2071$-$North. 
If the N$_{\rm NH_3}$/N$_{\rm CCS}$ ratio were dependent on 
the age of cloud cores only, B1$-$IRS would then be the youngest
source, and NGC2071$-$North the oldest one, but our ordering does not 
exactly conform to an evolutionary scheme. L1251A and NGC
2071-North have indeed been proposed to be the more evolved of these
six sources, but most of the CCS non-detections seem to be younger than
these two objects. On the other hand, we would expect GF9$-$2 to show
the lowest N$_{\rm NH_3}$/N$_{\rm CCS}$ ratio, given its classification
as a transitional object between prestellar and protostellar phase
\citep{Wie99}. These discrepancies are understandable if, as \citet{Suz92}
suggest, the evolutionary NH$_3$-CCS anticorrelation fades out when
using data from different sources.

The absence of anticorrelation among different clouds was explained by
\citet{Suz92} as the effect of the overlapping of individual evolutionary tracks for the chemistry of clouds with different H$_{2}$ column
 densities.
For this reason, the relationship between 
column densities of CCS and ammonia may not be a good quantitative
indicator of cloud evolution even assuming a common start time. 
This statement has also been suggested
by \citet{Benson98}, who did not find statistically
significant differences between star forming regions and starless
cores in their CCS column density.

Furthermore, even the abundance of CCS alone has been questioned as a good
quantitative tracer of cloud evolution. \citet{Lai00} suggested that 
 the abundance of CCS is not likely to trace the age of contracting cores very
 sensitively. Moreover \citet{deG05} proposed that CCS abundance could
 be enhanced through interaction with a molecular outflow, which suggests that
the local conditions and processes within each cloud could  also
influence the CCS emission. Being evident that CCS emission does
not depend on the age of the clouds only, we have searched for other
possible dependences on different parameters. 
To do this, we have compiled from the literature  
several parameters of the observed sources (luminosity, radio continuum and water
maser flux densities) and of their associated molecular outflows (degree of collimation, 
mass, dynamical age, mechanical
luminosity, kinetic energy, momentum, and momentum rate). For consistency,
we obtained the outflow parameters from single-dish observations of
the CO(2-1) line, when available. Otherwise, we made use of
interferometric and/or data from other CO transitions.

To check whether any of these parameters of sources and outflows
is related in any way with the presence of detectable CCS emission, 
we have tried to determine whether the parameters show different distributions
in the groups of sources with and without CCS emission. The
application of a Kolmogorov-Smirnov test did not find any significant
difference in distribution of the source or outflow parameters.

However, when we apply the same test to the
characteristics of the ammonia emission we measured, we find that
the distribution of the peak ammonia
intensity significantly depends on CCS detectability. 
Interestingly, we found no significant difference in the
distribution of integrated intensity (column density) of ammonia among
sources with and without CCS. 
In a more quantitative way, the mean peak intensity of NH$_3$ in the group of sources detected in
CCS is 3.3 K, while that in the group of undetected sources is 2.1 K.
We found that these two means are significantly different, 
using a T-test (and first applying a Lilliefors test to check
that the involved variables follow a normal distribution, which is a
prerequisite for the use of the T-test).

We have to be cautious in interpreting this result that sources with
detectable CCS emission show significantly brighter NH$_3$
emission. We have assumed in our calculations that the filling factor
of both emissions has a value of one, which is likely not to be true,
at least in the case of CCS, since interferometric maps in several
regions show its emission to be significantly clumpy \citep{Vel95,Lai00}. 
Therefore, the trend we have found could be just a distance effect since, 
if the molecular emission does not fill the Robledo beam, sources closer to us would
tend to be brighter in both CCS and NH$_3$.  However, we think
that the observed trend of higher peak NH$_3$ intensity in sources
with CCS is due to intrinsic characteristics of the
sources, rather than to a filling factor/distance effect. The most
important argument is that, if it were due to a distance bias, 
we would expect exactly the same trend for
the NH$_3$ integrated intensity but, as mentioned
above, we saw no significant difference for this quantity between
sources with and without detectable CCS emission. On the other hand,
it is not obvious that the subset of low-mass sources with CCS
emission are significantly closer to us than those without emission. 

A possible effect of different filling factors could be to hide
real correlations between source and cloud parameters and those of
the molecular line emission. However, we
do not think it likely that it is showing a spurious correlation in
this case. In any case, the impact of source distance on filling
factors is probably not very strong, if the whole extent of the
molecular emission is larger than the telescope beam. If this is the
case, the filling factor of a clumpy distribution, albeit lower than
1, would not change much with distance if clumps were more or less
uniformly distributed within the total spatial extent of the
emission. In the case of B1$-$IRS, CCS 
and NH$_{3}$ maps show emission $\simeq 3$ times more extended than
the Robledo beam ($\simeq 40''$). Ammonia maps of the sources sampled
here are also typically larger than this beam \citep{Jij99}. The lack of CCS maps in the 
star forming regions in which we detected this emission makes the
quantification of their filling factor difficult.

The fact that we find differences in peak intensity, but not in
integrated intensity can be explained if the significant
decrease of NH$_3$ peak intensity between sources with and without detectable
CCS emission is somewhat compensated by a corresponding increase in the NH$_3$
line width, resulting in small differences in integrated intensities.
This could be interpreted in evolutionary terms, since mass-loss phenomena 
related to active star forming processes would increase the turbulence of the
surrounding interstellar medium, thus broadening the line width as these
processes progress. Therefore, younger regions would tend to show
detectable CCS emission and narrower NH$_{3}$ lines. However, with our data
we cannot confirm such a trend for ammonia linewidths. 
A Kolmogorov-Smirnov test did not find any
significant difference in those linewidths between sources with and
without detected CCS. It may still be possible that there is a
difference, too small to be detected by the Kolmogorov-Smirnov test, 
but enough to blur the differences in integrated intensity between
sources with or without CCS.

It is also noticeable of interest to consider the different widths 
of the CCS and NH$_3$ lines
for the 6 sources detected in the former (see Tables \ref{tbl-CCS} and
\ref{tbl-NH3}). Ammonia thermal linewidths are a factor of $\simeq 1.8$
intrinsically broader than those of CCS,
due to the different masses of these molecules, but this thermal
broadening alone cannot explain the observed differences.
In the most extreme cases, L1448$-$IRS3
 and GF9$-$2, the NH$_3$ line is more than 3 times broader than the
 CCS one. This is
consistent with both lines tracing different regions of gas, with
different kinematics. The narrower CCS lines would
trace more quiescent gas, with less turbulence or velocity
gradients. B1-IRS is the source in which both line widths are more
similar, but \citet{deG05} shows that in this source, the
CCS-emitting gas shows evidence of interaction with the molecular
outflow, with clear velocity gradients, which makes CCS lines
wider. Apart from B1-IRS, the source with the most similar line widths
is L1448C, a source with an energetic molecular outflow. This could
suggest that the gas traced by CCS in this region may also be
significantly interacting with  the molecular outflow. In this context
the source GF9-2 would be the youngest source, in agreement with the 
transitional prestellar-protostellar classification made by  \citet{Wie99}.

We point out that these results must be considered carefully, since they 
are based on a small sample of regions showing CCS emission. The
lack of statistically significant results for the source and outflow
parameters indicates that CCS emission could be sensitive 
to the local temperature and density conditions of each
cloud, or to filling factor/distance effects. 
Moreover, we must take into account that the outflow and
source parameters were obtained from different instruments, 
and that on the other hand, the number of CCS detections is small, which
would mask possible correlations. 
We obviously need to widen this study to include
more star forming regions associated with CCS emission. Moreover, obtaining a
homogeneous set of observations of outflows and source parameters
(obtained with the same telescope for all sources)
would be useful for improving our statistical study. While such
studies will require a large amount of telescope time, it is not
possible to circumvent the fact that gas and dust in star forming
regions have very low surface brightness.

\subsection{Prospects for future interferometric observations}
There is not yet a theoretical chemical model
that includes the effects of a central protostar and explains how 
they affect CCS and ammonia formation and destruction.
High resolution observations of CCS and ammonia in 
star forming regions can provide the data against which such models could
be tested.

 Following the results obtained for the B1-IRS
region (that suggest an interaction of the CCS with a
molecular outflow that could enhance CCS via shocked-induced
chemistry, and show for the first time a spatial anticorrelation
between CCS and NH$_{3}$ at scales of $\simeq$5$''$) we propose that 
the six detections in this survey are 
good candidates for high angular resolution observations, to
compare with and to understand better the results obtained in B1-IRS.

All of these six sources are intrinsically interesting for detailed
interferometric observations. Also, L1448C and L1448-IRS3 are
especially appropriate for studying the suggested association between the
CCS and the molecular outflow, since this region hosts energetic and collimated
molecular outflows. Moreover, this region is a good environment for studying 
outflow-induced star formation. In GF9-2 it would be
interesting to study the CCS and the ammonia distribution
in a young object transitional between pre-stellar and protostellar Class 0 source.
NGC2071-North and L1251A are good prospective sites for making
high-resolution observations of CCS in regions classified as more
evolved (in which very few such observations have been made), or to  
try to reveal whether
there is a younger object driving their respective molecular outflows.

\section{Conclusions}
We present a single-dish survey of CCS and NH$_{3}$ emission at 1 cm, 
carried out with NASA's 70m antenna at Robledo,
 toward a sample of low- and intermediate-mass young star
forming regions that are known to harbor water maser emission. 
Our aim was to find the best candidates for a VLA high resolution
study of the kinematical and physical properties of young Class 0
objects, to search for the youngest protostars, and to determine the
relationship between CCS and NH$_{3}$ in star forming regions. Our general
conclusions are the following:
\begin{itemize}

\item We have detected 6 low-mass sources that show CCS emission, out of a
  sample of 40 star forming regions associated with water maser
  emission. Four of these (L1448C, L1448-IRS3, GF9-2, and L1251A) had
  not been previously detected in any CCS transition, another one
  (NGC 2071-North) has been detected for the first time with the CCS line
at 1 cm, while the other one
  (B1-IRS) had already been reported to show CCS emission in this
  and other 
  transitions. All our CCS detections also show ammonia emission.

\item From the CCS detection rate, and the duration of the water maser
  emission in low-mass star forming regions, we derive a lifetime of this
  molecule of $\simeq$(0.7$-$3) $\times$ 10$^{4}$ yr in these regions, after
  star formation started. 

\item Three of the six sources detected in CCS are catalogued as Class 0 protostars
  (L1448-IRS3, L1448C and B1-IRS), one could be a transitional object
  between pre-stellar and protostellar Class 0 stage
  (GF9-2), and the last two sources (NGC2071-North and L1251A) are
  catalogued as more evolved, a young T Tauri and a Class I source,
  respectively. 
  Since CCS is considered an early-time molecule, to explain the
  CCS detections in more evolved objects, we speculate
  with two possibilities: either the classification of NGC2071-North
  and L1251A should be revised, or the star formation activity and the
  physical properties of each cloud could influence in the production
  and destruction of the CCS molecules.

\item We did not find any statistically significant trend that may
  relate the presence of CCS emission with different parameters of
  the molecular outflows or the central sources of these star forming
  regions.

\item We found that the distribution and mean of the peak intensity of
  NH$_{3}$ in the group of sources detected in CCS are significantly
  different from those in the group of undetected ones, with the
  ammonia mean peak intensity higher in regions with CCS. However, no
  significant difference is found with respect to the integrated
  intensity of NH$_{3}$. Stronger NH$_{3}$ peak line intensities with
  indistinguishable integrated intensities suggests that the lines are
  narrower and the emitting regions less turbulent (i.e. younger).

\item The linewidths of the CCS and NH$_3$ lines are noticeably
  different, with CCS being 3 times narrower in some cases. This
  suggest that emission from these lines arises from gas with
  different kinematical properties within the telescope beam. The
  cases with more similar widths (B1-IRS, L1448C) may trace strong
  interactions between molecular outflows and the CCS-emitting gas.


\end{itemize}

\acknowledgments
JFG and IdG are partially supported by grant AYA2002-00376 of the
Spanish MEC (co-funded by FEDER funds). JFG is also supported by grant
AYA2005-08523-C03-03 of MEC. OS is partially supported by
grant AYA2003-09499 of MEC. EJ acknowledges partial support
by AYA2004-08260-C03-03 of MEC. The work by TBHK was done at the Jet
Propulsion Laboratory, California Institute of Technology, under
contract to the National Aeronautics and Space Administration. 
LFR acknowledges the support from DGAPA, UNAM, and CONACyT, Mexico. 
IdG acknowledges the support of a 
Calvo Rod\'es Fellowship from the Instituto Nacional de 
T\'ecnica Aeroespacial (INTA).
We are thankful to Jes\'us Calvo, Cristina Garc\'{\i}a, Esther Moll,
Pablo 
Perez, and the operators at the Madrid Deep Space Communication
Complex (MDSCC) for their help before and during the observations at
Robledo. 
This paper is based on observations taken during 
``host-country'' allocated time at Robledo de Chavela; this time is
managed by the Laboratorio de Astrof\'{\i}sica Espacial y F\'{\i}sica
Fundamental (LAEFF) of INTA, under agreement with National
Aeronautics and Space Administration/Ingenier\'{\i}a y Servicios
Aeroespaciales (NASA/INSA). We would also like to thank our anonymous
referee, whose valuable comments have greatly improved the quality of
this paper.

\begin{deluxetable}{lcccrcrl}
\tabletypesize{\scriptsize}
\tablecaption{Observed sources\label{tbl-source}}.
\tablehead{
\colhead{Source}&
\colhead{Alternative name}&
\colhead{Right Ascension\tablenotemark{a}}&
\colhead{Declination\tablenotemark{a}} &
\colhead{V$_{\rm LSR}$\tablenotemark{b}}&
\colhead{L$_{\rm bol}$\tablenotemark{c}}&
\colhead{D\tablenotemark{d}}&
\colhead{References\tablenotemark{e}}
\\
& &
\colhead{(J2000)}  & 
\colhead{(J2000)}&
\colhead{(km s$^{-1}$)}&
\colhead{(L$_{\rm \odot}$)}&
\colhead{(pc)}
}

\startdata

L1287\tablenotemark{f}              &IRAS 00338$+$6312        &00 36 47.5        &+63 29 02        &$-$18.0    &1000        &850      &11,67,42 \\
L1448$-$IRS3                        &IRAS 03225$+$3034        &03 25 36.4        &+30 45 20        &4.5        &10          &300      &12,5,8 \\
L1448C                              &LDN 1448-mm              &03 25 38.7        &+30 44 05        &4.5        &9           &300      &12,5,8 \\
RNO 15 FIR                          &IRAS 03245$+$3002        &03 27 39.0        &+30 12 59        &5.2        &16          &350      &12,63,60 \\
IRAS 2A                             &IRAS 03258+3104          &03 28 55.4        &+31 14 35        &7.0        &43          &350      &24,41,23 \\
HH 6                                &IRAS 7                   &03 29 11.2        &+31 18 31        &7.0        &18          &350      &33,19,39 \\
B1$-$IRS                            &IRAS 03301+3057          &03 33 16.3        &+31 07 51        &6.3        &2.8         &350      &24,36 \\
T TAU$-$South                       &IRAS 04190+1924          &04 21 59.4        &+19 32 06        &8.2        &10          &160      &12,46,10 \\
L1534B                              &IRAS 04361+2547          &04 39 13.9        &+25 53 21        &6.2        &3.8         &140      &12,35,8 \\
L1641$-$North\tablenotemark{f}      &IRAS 05338$-$0624        &05 36 18.7        &$-$06 22 09      &7.4        &120         &500      &12,61,22 \\
HH 1                                &IRAS 05339$-$0647        &05 36 19.1        &$-$06 45 01      &9.3        &50          &500      &16,57,32 \\
AFGL 5157\tablenotemark{f}          &IRAS 05345+3157          &05 37 47.8        &+31 59 24        &$-$18.0    &5000        &1800     &34,48,52 \\
Haro4$-$255                         &IRAS 05369$-$0728        &05 39 22.3        &$-$07 26 45      &4.8        &13          &480      &12,2,21 \\
L1641$-$S3                          &IRAS 05375$-$0731        &05 39 56.0        &$-$07 30 18      &5.1        &70          &450      &12,61,53 \\
NGC 2024 FIR 5                      &Orion B                  &05 41 44.5        &$-$01 55 43      &11.0       &$\ga$10     &450      &24,29,1 \\
HH 212                              &IRAS 05413$-$0104        &05 43 51.1        &$-$01 03 01      &1.7        &14          &500      &12,30,68 \\
B35A                                &IRAS 05417+0907          &05 44 29.8        &+09 08 54        &11.8       &15          &460      &12,7,6 \\
HH 19$-$27                          &NGC 2068 H$_{2}$O maser  &05 46 31.2        &$-$00 02 35      &9.9        &1.7         &400      &3,48,25 \\
NGC 2071\tablenotemark{f}           &IRAS 05445$+$0020        &05 47 04.8        &+00 21 43        &9.4        &750         &500      &56,64,31 \\
NGC 2071$-$North                    &IRAS 05451+0037          &05 47 42.3        &+00 38 40        &9.0        &40          &500      &12,64,38 \\
IRAS 06291+0421\tablenotemark{f}         &                    &06 31 48.1        &+04 19 31        &13.7       &1882        &1600     &9,66,65 \\
NGC 2264 IRS\tablenotemark{f}       &IRAS 06384+0932          &06 41 10.3        &+09 29 19        &8.7        &2300        &800      &22,64,43 \\
IRAS 06584$-$0852\tablenotemark{f}       &                    &07 00 51.5        &$-$08 56 29      &40.5       &5670        &4480     &49,66,45 \\
CB 54                               &IRAS 07020-1618          &07 04 21.2        &$-$16 23 15      &19.5       &55          &600      &27,13,58 \\ 
L260                                &IRAS 16442-0930          &16 46 58.6        &$-$09 35 23      &3.5        &0.97        &160      &22,28,47 \\
IRAS 18265+0028\tablenotemark{f}         &                    &18 29 05.8        &+00 30 36        &5.3        &347         &440      &14,49 \\
Serpens FIRS 1                      &IRAS 18273+0113          &18 29 49.8        &+01 15 21        &8.0        &46          &310      &12,20,37 \\
LDN 723$-$mm                        &IRAS 19156+1906          &19 17 53.9        &+19 12 25        &10.5       &3           &300      &24,26,17 \\
IRAS 20050+2720\tablenotemark{f}    &IRAS 20050+2720 MMS 1    &20 07 06.7        &+27 28 53        &6.0        &260         &700      &9,4,61 \\
S106 FIR\tablenotemark{f}           &IRAS 20255+3712          &20 27 25.5        &+37 22 49        &$-$1.0     &$<$1000     &600      &24,63,50 \\
L1157-mm                            &IRAS 20386+6751          &20 39 06.5        &+68 02 13        &2.7        &11          &440      &12,44,18 \\
GF9$-$2                                   &                   &20 51 30.1        &+60 18 39        &$-$2.7     &0.3         &200      &24,59 \\
B361                                &IRAS 21106+4712          &21 12 26.1        &+47 24 24        &2.7        &4.7         &350      &12,28,6 \\
CB 232                              &IRAS 21352+4307          &21 37 11.3        &+43 20 36        &12.6       &3.9         &350      &27,13,15 \\
IC 1369N\tablenotemark{f}           &IRAS 21391+5802          &21 40 42.4        &+58 16 10        &0.2        &500         &750      &9,62,54 \\
L1204A\tablenotemark{f}            &IRAS 22198+6336           &22 21 27.6        &+63 51 42        &$-$10.5    &367         &900      &24,63,22 \\
L1204B                              &IRAS 22199+6322          &22 21 33.3        &+63 37 21        &$-$10.3    &52          &900      &14,55,22 \\
L1251A                              &IRAS 22343+7501          &22 35 24.3        &+75 17 06        &$-$4.0     &27          &300      &12,28,51 \\
L1251B                              &IRAS 22376+7455          &22 38 47.1        &+75 11 29        &$-$4.0     &14          &300      &12,51 \\
Cepheus E                           &IRAS 23011+6126          &23 03 13.1        &+61 42 26        &$-$10.4    &50          &730      &24,64,40 \\                              

\enddata 
\tablenotetext{a}{Observed position, coincident with reported water maser position. Units of right ascension are hours, minutes, and  seconds. Units of declination 
are degrees, arcminutes, and arcseconds}                    
\tablenotetext{b}{Velocity of the cloud with respect to the local standard of rest}.
\tablenotetext{c}{Bolometric luminosity of the source}.
\tablenotetext{d}{Distance to the source}.
\tablenotetext{e}{REFERENCES:(1)\citealt{And00}, (2)\citealt{Ang89}, (3)\citealt{Ang96},
  (4)\citealt{Bac95}, (5)\citealt{Bac90}, (6)\citealt{Bei86}, (7)\citealt{Ben89},
  (8)\citealt{Bon96}, (9)\citealt{Bra94}, (10)\citealt{Cab92}, (11)\citealt{Ces88}, (12)\citealt{Cla96},
 (13)\citealt{Cle88}, (14)\citealt{Cod95}, (15)\citealt{Cod97}, (16)\citealt{Com90}, (17)\citealt{Dav87}, (18)\citealt{Dav95},  
(19)\citealt{Edw83}, (20)\citealt{Eir92}, (21)\citealt{Eva86}, (22)\citealt{Fel92},
(23)\citealt{Fro05}, (24)\citealt{Fur03}, (25)\citealt{Gib00}, (26)\citealt{Gol84}, (27)G{\'o}mez et al. in preparation,
 (28)\citealt{Goo93}, (29)\citealt{Gra93}, (30)\citealt{Har93}, (31)\citealt{Har79}, (32)\citealt{Har86}, (33)\citealt{Hen86}, (34)\citealt{Hen92}, (35)\citealt{Her87}, (36)\citealt{Hir97}, (37)\citealt{Hur96}, (38)\citealt{Iwa88}, (39)\citealt{Jen87}, (40)\citealt{Lad97}, (41)\citealt{Lan96}, (42)\citealt{Lor00}, (43)\citealt{Mar89}, (44)\citealt{Mik92}, (45)\citealt{Mol00}, (46)\citealt{Mom96}, (47)\citealt{Mye87}, (48)\citealt{Pas91}, (49)\citealt{Per94}, (50)\citealt{Ric93}, (51)\citealt{Sat94}, 
(52)\citealt{Sne88}, (53)\citealt{Sta00}, (54)\citealt{Sug89}, (55)\citealt{Taf93},
(56)\citealt{Tof95}, (57)\citealt{Tor93}, (58)\citealt{Wan95}, 
(59)\citealt{Wie99}, (60)\citealt{Wil94conf}, (61)\citealt{Wil89}, (62)\citealt{Wou89}, (63)\citealt{Wou93}, 
(64)\citealt{Wou88}, (65)\citealt{Wu04}, (66)\citealt{Wu01}, (67)\citealt{Yan95}, (68)\citealt{Zin92}}
\tablenotetext{f}{Intermediate-luminosity (L$_{bol}$ $>$ 100
  L$_{\odot}$) sources.}

\end{deluxetable}

\begin{deluxetable}{lccccccccccc}
\tabletypesize{\scriptsize}
\tablecaption{CCS (2$_{1}$-1$_{0}$) line parameters \tablenotemark{a}\label{tbl-CCS}}
\tablewidth{0pt}
\tablehead{
\colhead{Source}&
\colhead{T$_{\rm MB}$ \tablenotemark{b}}&
\colhead{V$_{\rm LSR}$ \tablenotemark{c}}&
\colhead{$\Delta$v \tablenotemark{c}}&
\colhead{${\rm \int{T_{MB}dv}}$ \tablenotemark{d}}&
\colhead{N$_{\rm CCS}$ \tablenotemark{e}}
\\
&
\colhead{(K)}& 
\colhead{(km s$^{-1}$)}&
\colhead{(km s$^{-1}$)}&
\colhead{(K km s$^{-1}$ )}& 
\colhead{(10$^{12}$ cm$^{-2}$)}& 

}
\startdata

L1287                &$<$0.3                         &\nodata           &\nodata          &$<$0.07          &$<$1.9 \\               
L1448$-$IRS3         &0.32$\pm$0.11                  &4.68$\pm$0.05     &0.36$\pm$0.11    &0.09$\pm$0.03    &2.4$\pm$0.8 \\         
L1448C               &0.46$\pm$0.07                  &4.65$\pm$0.03     &0.58$\pm$0.07    &0.305$\pm$0.024  &8.1$\pm$0.6 \\         
RNO 15 FIR           &$<$0.15                        &\nodata           &\nodata          &$<$0.03          &$<$0.8 \\              
IRAS 2A              &$<$0.21                         &\nodata           &\nodata         &$<$0.05          &$<$1.3 \\              
HH 6                 &$<$0.24                         &\nodata           &\nodata         &$<$0.05          &$<$1.3 \\              
B1$-$IRS             &0.70$\pm$0.13                  &6.72$\pm$0.04     &0.89$\pm$0.10    &0.57$\pm$0.05    &15.1$\pm$1.3 \\        
T TAU$-$South        &$<$0.22                         &\nodata           &\nodata         &$<$0.05          &$<$1.3 \\              
L1534B               &$<$0.20                         &\nodata           &\nodata         &$<$0.04          &$<$1.1 \\              
L1641$-$North\tablenotemark{f}        &$<$0.20       &\nodata           &\nodata          &$<$0.05          &$<$1.3 \\              
HH 1\tablenotemark{f}                 &$<$0.24       &\nodata           &\nodata          &$<$0.06          &$<$1.6 \\              
AFGL 5157            &$<$0.3                         &\nodata           &\nodata          &$<$0.07          &$<$1.9 \\              
Haro4$-$255          &$<$0.3                         &\nodata           &\nodata          &$<$0.07          &$<$1.9 \\              
L1641$-$S3           &$<$0.22                         &\nodata           &\nodata         &$<$0.05          &$<$1.3 \\              
NGC 2024 FIR 5       &$<$0.20                         &\nodata           &\nodata         &$<$0.04          &$<$1.1 \\              
HH 212               &$<$0.3                         &\nodata           &\nodata          &$<$0.07          &$<$1.9 \\              
B35A                 &$<$0.23                         &\nodata           &\nodata         &$<$0.05          &$<$1.3 \\              
HH 19$-$27           &$<$0.3                         &\nodata           &\nodata          &$<$0.07          &$<$1.9 \\              
NGC 2071\tablenotemark{f}             &$<$0.24      &\nodata           &\nodata           &$<$0.06          &$<$1.6 \\              
NGC 2071$-$North     &0.41$\pm$0.12                  &8.54$\pm$0.04&0.38$\pm$0.08         &0.14$\pm$0.03    &3.7$\pm$0.8 \\          
IRAS 06291+0421      &$<$0.3                         &\nodata           &\nodata          &$<$0.07          &$<$1.9 \\              
NGC 2264 IRS         &$<$0.20                         &\nodata           &\nodata         &$<$0.04          &$<$1.1 \\              
IRAS 06584$-$0852    &$<$0.23                         &\nodata           &\nodata         &$<$0.05          &$<$1.3 \\              
CB 54                &$<$0.22                         &\nodata           &\nodata         &$<$0.05          &$<$1.3 \\              
L260                 &$<$0.5                         &\nodata           &\nodata          &$<$0.11          &$<$3 \\                
IRAS 18265+0028      &$<$0.20                         &\nodata           &\nodata         &$<$0.04          &$<$1.1 \\              
Serpens FIRS 1       &$<$0.20                         &\nodata           &\nodata         &$<$0.04          &$<$1.1 \\              
LDN 723$-$mm         &$<$0.3                         &\nodata           &\nodata          &$<$0.07          &$<$1.9 \\              
IRAS 20050+2720      &$<$0.3                         &\nodata           &\nodata          &$<$0.07          &$<$1.9 \\              
S106 FIR             &$<$0.15                         &\nodata           &\nodata         &$<$0.03          &$<$0.8 \\              
L1157$-$mm\tablenotemark{f}           &$<$0.18       &\nodata           &\nodata          &$<$0.05          &$<$1.3 \\              
GF9$-$2\tablenotemark{f}     &0.90$\pm$0.16        &$-$2.415$\pm$0.022 &0.24$\pm$0.05     &0.25$\pm$0.05    &6.6$\pm$1.3 \\         
B361                 &$<$0.16                         &\nodata           &\nodata         &$<$0.04          &$<$1.1 \\              
CB 232               &$<$0.24                         &\nodata           &\nodata         &$<$0.05          &$<$1.3 \\              
IC 1369N             &$<$0.3                         &\nodata           &\nodata          &$<$0.07          &$<$1.9 \\              
L1204A               &$<$0.24                         &\nodata           &\nodata         &$<$0.05          &$<$1.3 \\              
L1204B               &$<$0.3                         &\nodata           &\nodata          &$<$0.07          &$<$1.9 \\              
L1251A\tablenotemark{f}    &0.59$\pm$0.11  &$-$4.91$\pm$0.04  &0.47$\pm$0.09              &0.28$\pm$0.04    &7.4$\pm$1.1 \\         
L1251B\tablenotemark{f}     &$<$0.3                  &\nodata           &\nodata          &$<$0.08          &$<$2.1 \\              
Cepheus E            &$<$0.3                           &\nodata           &\nodata        &$<$0.07          &$<$1.9 \\

\enddata

 \tablenotetext{a}{Uncertainties in this table are $2\sigma$ for detections. For nondetections, upper limits are $3\sigma$.}
\tablenotetext{b}{Main beam brightness temperature.}
\tablenotetext{c}{Central V$_{\rm LSR}$ and line width obtained from a
  Gaussian fit to the  CCS line. Uncertainties represent 
the error in the Gaussian fit.}
\tablenotetext{d}{Integrated intensity. Upper limits were calculated
  assuming a  width equal to the mean value of velocity width for
  detections ($\Delta$v $\simeq$ 0.5 km s$^{-1}$).}
\tablenotetext{e}{CCS column density obtained from $N_{\rm {mol}} = \frac
  {8 \pi \nu^{3}}{c^{3} g_{j} A_{ji}} \frac{Q(T_{\rm{rot}}) \int {T_MB dv}}{[B_\nu(T_{\rm {ex}}) - B_\nu(T_{\rm
      {bg}})]}\frac{\exp({E_{j}/kT_{\rm{rot}}})}{\exp(h\nu/kT_{\rm ex})-1}$
  (optically thin approximation), $\nu$ is the frequency of the transition, $Q$
is the partition function, 
E$_{j}$ is the energy of the upper state (1.61 K; \citealt{Wol97}),
 $T_{\rm rot}$ is the rotational temperature, $g_{j}$ is the statistical weight of the upper rotational level.
$A_{ji}$ is the Einstein coefficient for the  CCS(2$_1$-1$_0$)
transition (4.33 $\times$ 10$^{-7}$s$^{-1}$;\citealt{Wol97}), $B(T)$
is the intensity of black-body radiation at temperature $T$, 
and $T_{\rm ex}$ is the excitation temperature. Since we only observed
one CCS
transition, we could not obtain a reliable estimate of $T_{\rm exc}$ and
$T_{\rm rot}$; therefore, we assumed $T_{\rm rot}$=$T_{\rm exc}$= 5 K, the mean value
assigned by \citet{Suz92} for a large sample of young sources. We assumed
these values even for the intermediate-mass sources, since, for instance, an
increase of 4 K in both $T_{\rm exc}$ and $T_{\rm rot}$ causes a change in the CCS column
density of only $\simeq 45$\%.} 
\tablenotetext{f}{Sources observed using the 384 channel spectrometer, with a velocity
  resolution of 0.07 km s$^{-1}$. These sources were smoothed to
  a final velocity resolution of $\simeq$ 0.14 km s$^{-1}$. 
   The rest of CCS targets (observed with the 256 channel spectrometer  with a velocity
  resolution of 0.05 km s$^{-1}$) were
  smoothed to a resolution of 0.10 km s$^{-1}$.}
\end{deluxetable}

\begin{deluxetable}{lccccccccccc}
\tabletypesize{\scriptsize}
\tablecaption{NH$_{3}$(1,1) line parameters \tablenotemark{a}\label{tbl-NH3}}
\tablewidth{0pt}
\tablehead{
\colhead{Source}&
\colhead{T$_{\rm MB}$ \tablenotemark{b}}&
\colhead{V$_{\rm LSR}$ \tablenotemark{c}}&
\colhead{$\Delta$v \tablenotemark{c}}&
\colhead{${\rm \int{T_{MB}dv}}$ \tablenotemark{d}}&
\colhead{N$_{\rm NH3}$ \tablenotemark{e}}
\\
&
\colhead{(K)}& 
\colhead{(km s$^{-1}$)}&
\colhead{(km s$^{-1}$)}&
\colhead{(K km s$^{-1}$ )}& 
\colhead{(10$^{15}$ cm$^{-2}$)}
}
\startdata

L1287                  &3.9$\pm$0.3      &-17.73$\pm$0.12          &2.2$\pm$0.3             &9.3$\pm$0.5    &2.76$\pm$0.15      \\
L1448$-$IRS3           &5.7$\pm$0.3      &  4.22$\pm$0.07          &1.17$\pm$0.16           &7.7$\pm$0.4    &1.52$\pm$0.08     \\
L1448C                 &4.97$\pm$0.24    &  4.60$\pm$0.03          &1.28$\pm$0.14           &6.8$\pm$0.3    &1.00$\pm$0.04     \\
RNO 15 FIR             &2.73$\pm$0.21    &  4.50$\pm$0.06          &1.00$\pm$0.14           &3.2$\pm$0.3    &1.33$\pm$0.12      \\
IRAS 2A                &1.70$\pm$0.13    &  7.27$\pm$0.09          &1.69$\pm$0.19           &3.15$\pm$0.17  &5.0$\pm$0.3       \\
HH 6                   &1.8$\pm$0.3      &  8.27$\pm$0.10          &1.1$\pm$0.3             &2.3$\pm$0.5    &2.4$\pm$0.5       \\
B1$-$IRS               &4.04$\pm$0.24    &  5.95$\pm$0.06          &1.06$\pm$0.14           &5.2$\pm$0.4    &1.79$\pm$0.14      \\
T TAU$-$South          &$<$0.3           &\nodata                  &\nodata                 &$<$0.3         &$<$0.09           \\
L1534B                 &$<$0.5           &\nodata                  &\nodata                 &$<$0.4         &$<$0.12           \\        
L1641$-$North          &3.68$\pm$0.21    &  7.11$\pm$0.04          &1.27$\pm$0.18           &5.2$\pm$0.3    &1.25$\pm$0.07     \\
HH 1                   &0.83$\pm$0.22    &  9.02$\pm$0.07          &2.1$\pm$0.5             &1.8$\pm$0.3    &27$\pm$5          \\
AFGL 5157\tablenotemark{f}              &$<$0.3           &\nodata                  &\nodata                 &$<$0.3         &$<$0.09           \\
Haro4$-$255            &1.7$\pm$0.3      &  4.44$\pm$0.10          &1.07$\pm$0.24           &2.0$\pm$0.4    &3.2$\pm$0.7       \\
L1641$-$S3             &2.6$\pm$0.4      &  5.28$\pm$0.12          &1.5$\pm$0.3             &4.0$\pm$0.5    &3.4$\pm$0.4       \\
NGC 2024 FIR 5         &1.94$\pm$0.22    & 11.12$\pm$0.13          &2.5$\pm$0.3             &4.8$\pm$0.4    &1.17$\pm$0.10      \\
HH 212                 &1.16$\pm$0.21    &  1.65$\pm$0.06          &0.8$\pm$0.3             &1.4$\pm$0.3    &11.2$\pm$2.4        \\
B35A                   &2.25$\pm$0.21    & 11.88$\pm$0.06          &1.04$\pm$0.17           &2.7$\pm$0.3    &2.05$\pm$0.22      \\
HH 19$-$27             &1.57$\pm$0.13    & 10.24$\pm$0.06          &1.12$\pm$0.16           &2.34$\pm$0.22  &4.4$\pm$0.4       \\          
NGC 2071               &3.12$\pm$0.24    &  8.94$\pm$0.07          &1.44$\pm$0.18           &5.0$\pm$0.4    &0.41$\pm$0.03     \\
NGC 2071$-$North\tablenotemark{f}       &1.02$\pm$0.07    &  8.37$\pm$0.11          &1.03$\pm$0.26           &0.94$\pm$0.13  &4.5$\pm$0.6       \\         
IRAS 06291+0421        &$<$0.3           &\nodata                  &\nodata                 &$<$0.3         &$<$0.09           \\
NGC 2264 IRS           &5.09$\pm$0.23    &  8.22$\pm$0.12          &2.7$\pm$0.3             &13.2$\pm$0.4   &1.42$\pm$0.04     \\
IRAS 06584$-$0852      &$<$0.3           &\nodata                  &\nodata                 &$<$0.3         &$<$0.09           \\
CB 54                  &0.98$\pm$0.13    & 19.55$\pm$0.09          &1.3$\pm$0.3             &1.40$\pm$0.21  &0.140$\pm$0.021    \\   
LDN 723$-$mm           &1.63$\pm$0.18    & 10.89$\pm$0.06          &0.88$\pm$0.15           &1.63$\pm$0.23  &0.67$\pm$0.09     \\
IRAS 20050+2720        &2.27$\pm$0.11    &  5.88$\pm$0.11          &2.5$\pm$0.3             &6.19$\pm$0.20  &3.80$\pm$0.12     \\
S106 FIR               &0.78$\pm$0.19    & -1.57$\pm$0.16          &1.4$\pm$0.4             &1.0$\pm$0.3    &0.56$\pm$0.18      \\        
L1157$-$mm             &2.24$\pm$0.17    &  2.64$\pm$0.06          &0.84$\pm$0.12           &2.00$\pm$0.21  &1.12$\pm$0.12     \\        
GF9$-$2                &3.32$\pm$0.18    & -2.49$\pm$0.06          &0.80$\pm$0.11           &2.71$\pm$0.22  &1.36$\pm$0.11      \\
B361                   &2.3$\pm$0.3      &  2.76$\pm$0.10          &1.5$\pm$0.3             &3.5$\pm$0.4    &0.28$\pm$0.03     \\
CB 232                 &1.00$\pm$0.11    & 12.39$\pm$0.07          &0.99$\pm$0.15           &1.10$\pm$0.15  &2.7$\pm$0.4       \\
IC 1369N               &1.79$\pm$0.21    &  0.50$\pm$0.11          &2.1$\pm$0.3             &4.2$\pm$0.4    &1.79$\pm$0.17      \\
L1204A                 &1.78$\pm$0.19    &-11.00$\pm$0.09          &1.44$\pm$0.19           &2.9$\pm$0.3    &4.4$\pm$0.5       \\
L1204B                 &1.49$\pm$0.22    &-10.56$\pm$0.10          &1.39$\pm$0.22           &2.2$\pm$0.3    &2.9$\pm$0.4       \\ 
L1251A                 &0.75$\pm$0.13    & -5.33$\pm$0.12          &1.4$\pm$0.3             &1.11$\pm$0.18  &6.5$\pm$1.1       \\
Cepheus E              &1.03$\pm$0.21    &-11.05$\pm$0.11          &1.01$\pm$0.24           &1.0$\pm$0.3    &2.2$\pm$0.7       \\

\enddata                                                                                                                                                      
\tablenotetext{a}{Uncertainties in this table are $2\sigma$ for detections. For nondetections, upper limits are $3\sigma$.}
\tablenotetext{b}{Main beam brightness temperature of the main
  hyperfine component.}
\tablenotetext{c}{Central V$_{\rm LSR}$ and line width obtained from a Gaussian fit to the main line of the NH$_3$(1,1) transition. Uncertainties represent 
the error in the Gaussian fit.}
\tablenotetext{d}{Integrated intensity of the main  hyperfine component. Upper limits were calculated assuming a width equal to
  the mean value of velocity width for detections ($\Delta$v $\simeq$ 1.4  km s$^{-1}$).}
\tablenotetext{e}{Column density obtained from $N_{\rm {mol}} = \frac
  {16 [1+\exp(h\nu/kT_{\rm ex})] \pi \nu^{3}}{c^{3} g_{j} A_{ji}} \frac{Q(T_{\rm{rot}}) \int {T_MB dv}}{[B_\nu(T_{\rm {ex}}) - B_\nu(T_{\rm
      {bg}})]}\frac{\exp({E_{j}/kT_{\rm{rot}}})}{\exp(h\nu/kT_{ex})-1}$
  (optically thin approximation). 
$E_{j}$ is the energy of the $J,K=1,1$ rotational level  (23.4 K;
\citealt{Ho83}),  
$g_{j}$ is the statistical weight of the upper sublevel involved in
the inversion transition, 
$A_{ji}$ is the Einstein coefficient of the NH$_3$(1,1) transition 
(1.67 $\times$
10$^{-7}$s$^{-1}$;  \citealt{Ho83}). 
We calculated the optical depth and the excitation temperature from the relationship between hyperfine components for each
  source. When no satellite hyperfine line was detected, we adopted
$T_{\rm ex}=7.8$ K,
the mean $T_{\rm ex}$ of all the sources. 
We assumed $T_{\rm rot} = T_{\rm ex}$.
} 
\tablenotetext{f}{Sources observed with a velocity resolution of 0.13 km~s$^{-1}$. These sources were smoothed to a final velocity resolution of 0.5 km~s$^{-1}$. The rest of the ammonia targets were observed with a velocity resolution of 0.5 km~s$^{-1}$.}
\end{deluxetable}

\end{document}